# Machine Learning-Aided Discovery of Superionic Solid-State Electrolyte for Li-Ion Batteries

Seungpyo Kang, Minseon Kim, and Kyoungmin Min [*]

School of Mechanical Engineering, Soongsil University, 369 Sangdo-ro, Dongjak-gu, Seoul 06978, Republic of Korea

## ABSTRACT

Li-Ion Solid-State Electrolytes (Li-SSEs) are a promising solution that resolves the critical issues of conventional Li-Ion Batteries (LIBs) such as poor ionic conductivity, interfacial instability, and dendrites growth. In this study, a platform consisting of a high-throughput screening and a machine-learning surrogate model for discovering superionic Li-SSEs among 20,237 Li-containing materials is developed. For the training database, the ionic conductivity of Na SuperIonic CONductor (NASICON) and Li SuperIonic CONductor (LISICON) type SSEs are obtained from the previous literature. Then, the chemical descriptor (CD) and additional structural properties are used as machine-readable features. Li-SSE candidates are selected through the screening criteria, and the prediction on the ionic conductivity of those is followed. Then, to reduce uncertainty in the surrogate model, the ensemble method by considering the best-performing two models is employed, whose mean prediction accuracy is 0.843 and 0.829, respectively. Furthermore, first-principles calculations are conducted for confirming the ionic conductivity of the strong candidates. Finally, six potential superionic Li-SSEs that have not previously been investigated are proposed. We believe that the constructed platform can accelerate the search for Li-SSEs with high ionic conductivity at minimum cost.

## KEYWORDS

[*]Corresponding author: kmin.min@ssu.ac.kr (K. Min)

**INTRODUCTION**

LIBs are widely implemented in electric vehicles, energy storage systems, and portable electronic devices because of their high energy density.[1,2] Most electrolytes used in commercial batteries consist of liquids in which Li salts and organic solvents.[3–5] These electrolytes are essential to increase the mobility of Li ions, and thus a key factor in battery performance.[2] However, the use of liquid electrolytes can cause a critical safety issue because of their flammability, potentially leading to ignition of solvents.[6,7] Additionally, Li dendrite growth in LIBs causes to reduce the structural stability and triggering thermal runway and explosion by an internal short circuit.[8–13] Thus, the use of solid-state electrolytes (SSEs) has emerged as a promising solution and active research is underway.[14–20]

Employing SSEs can greatly improve safety because of their nonflammability and enhance the overall energy density in LIBs. In addition, the high shear modulus of such electrolytes inhibits the growth of Li-dendrites.[12,13,17] However, compared to liquid-type electrolytes, their relatively low ionic conductivity and high interfacial resistance between electrode and SSE are a major bottleneck in the commercialization of solid-state Li-ion batteries.[21–25] Ideal SSEs should satisfy the following conditions. (1) high ionic conductivity, (2) wide electrochemical stability windows, (3) insignificant electronic conductivity, (4) robust chemical stability, and (5) low manufacturing cost.[7,26] Among these, moderate-high ionic conductivity (ionic conductivity ($\sigma = 10^{-4}$ S/cm) at room temperature (RT) is essential for their use at a wide range of operating temperatures. However, the application of SSE is currently limited, because SSEs achieve liquid electrolyte-like conductivity ($\sigma = 10^{-2}$ S/cm) at only around 50-80 ℃.[27] Over the past few decades, there are ongoing studies for discovering new SSEs, but only a handful of candidates with ionic conductivity comparable to that of a liquid electrolyte were identified such as LISICON type $Li_{10}GeP_2S_{12}$, argyrodite type $Li_6PS_5Br$, and garnet type $Li_{6.55}La_3Zr_2Ga_{0.15}O_{12}$.[27] In this respect, a large number of candidates have not yet been examined, thus materials with high ionic conductivity at RT could exist among numerous Li-containing solids.

The method of experimentally obtaining and computationally calculating all properties of a wide range of materials is inefficient and impossible because it requires a lot of time and cost.

For this reason, machine learning-based methodologies are actively utilized in materials science and related fields because they guarantee high accuracy and speed.[28–34] Representative examples using machine learning for energy-relevant fields are as follows; (1) the electrochemical properties prediction of electrode materials,[35] (2) prediction of shear and bulk modulus for Li-SSEs,[36] and (3) design of superionic Li-ion conductors based on optimization of materials compositions.[37] In this respect, the implementation of machine learning algorithms is expected to significantly reduce the overall time and cost for materials screening to discover novel Li-SSEs with ideal ionic conductivity.

In this study, we built a platform consisting of high-throughput screening and a machine learning model to classify the ionic conductivity of Li-SSE candidates in RT. For constructing the training database, ionic conductivity values of NASICON and LISICON were obtained from the previous references.[28,38] For the testing database, over 20,000 Li-containing materials were obtained from the Materials Project (MP) library. For optimal performance of the surrogate model, an ensemble method of two algorithms was applied. Then, the ionic conductivity class for screened Li-SSE candidates is predicted to identify the potential super-ionicity. Finally, for validating the surrogate model and candidates, density functional theory (DFT) molecular dynamics calculations were performed. As a result, we were able to propose new superionic Li-SSE candidates that have not been previously studied.

**METHODS**

**Figure 1(a)** shows all the processes performed in this study. The first process involves an initial database construction; the ionic conductivity of NASICON and LISICON structures is obtained from the previous literature.[28,38] As superionic SSE candidate materials, Li-containing materials (20,237) are extracted from MP database.[39,40] After constructing the initial database, Li-SSE candidates screening process is performed. Through the screening process, 468 Li-SSE candidates with a high possibility of being superionic conductors are derived. Then a classification model is constructed by ensembling two models. The developed surrogate model is used to predict the ionic conductivity class of 468 screened Li-SSE candidates at RT. For 42 Li-SSEs expected to be superionic, we performed DFT calculations to obtain the ionic conductivity values of these materials. Through this process, the performance of the surrogate model and the ionic conductivity of suggested materials can be verified. Finally, we propose verified superionic Li-SSE candidates that have not yet been researched so far.

**Screening Process**

To select Li-SSE candidates with a high likelihood of superionic nature, a screening process is performed. As shown in **Figure 1(b)**, the screening process considered multiple steps to remove unsatisfactory materials as SSEs;[7,26] (1) stability against reduction at Li metal anode, (2) electronic conductivity, (3) structural stability, and (4) stability against oxidation at cathode. First, the structures with a transition metal (TM) atom are particularly susceptible to reaction with Li, as various stable oxidation states exist in TMs. It is known that materials containing TMs do not maintain stability against Li metal.[41] Second, the electronic conduction across the electrolyte must be minimized in batteries. In this respect, a large band gap (≥1.0 eV) must be guaranteed for minimal electronic conductivity. Third, for the prevention of decomposition reaction, the energy above the convex hull ($E_{hull}$) of SSE candidates should be 0 eV/atom. Finally, stability against oxidation at the cathode is related to an oxidative voltage ($V_{ox}$), we contain materials whose $V_{ox}$ > 3.7 V. Following these criteria, the number of Li-SSE candidates is selected as follows; (1) 3,190 (2) 1,859, (3) 567, and finally, (4) 468 materials. 468 Li-SSE candidates are expected to exhibit great potential as SSEs.

**Machine Learning Details**

The machine learning process for predicting superionic Li-SSE candidates is shown in **Figure 1(c)**. For constructing a surrogate model, we considered the composition of the features and the size of the database. As the feature, previously developed CD and additional structural properties are considered. CD consists of 145 features that represent stoichiometric, elemental properties, and so on (**Table S1, Supporting Information (SI)**).[42] Additional structural properties consist of lattice parameters (a, c, and a/c), volume, and space group number. For the cross-validation, we considered mean accuracy for 100 different train-validation set compositions from nine different algorithms such as Naive Bayes (NB), Logistic Regression (LR), Decision Tree (DT), Stochastic Gradient Descent (SGD), K-Nearest Neighbor (KNN), RandomForest (RF), Gradient Boosting (GB), Light Gradient Boosting Machine (LGBM), and Extreme Gradient Boosting (XGB) classifier. Hyperparameter optimization is carried out for each algorithm.

**Ab initio Molecular Dynamics simulations**

To compute the ionic conductivity values, we apply an *ab initio* molecular dynamics (AIMD) scheme in Vienna *ab initio* simulation package (VASP).[43,44] For the exchange-correlation functional, the Perdew–Burke–Ernzerhof generalized gradient approximation is applied.[45] The plane-wave cutoff is set to 400 eV. For k-point grid generation, the gamma-centered 1 × 1 × 1 is used. The time-step of 2.0 fs is applied. AIMD simulations are performed as follows: (1) The structure is equilibrated at each temperature (600K, 800K, 1000K, and 1200K) for 3,000 time step (6 ps) in the NVT ensemble with Nose-Hoover thermostat. (2) For describing the diffusion, NVT ensemble is further applied for 10,000 time steps (20 ps). (3) The diffusion coefficient is obtained based on the mean squared displacement and the activation energies are determined from the Arrhenius plots with respect to the diffusivity. The RT ionic conductivity is obtained by extrapolating the diffusion coefficient vs. temperature under considered temperature region. The further details can be found in ref.[46]

**RESULTS AND DISCUSSIONS**

**Database Construction**

To construct a training database for developing the surrogate model, we obtained the information on the ionic conductivity of NASICON (209) and LISICON (55) SSEs in the previous literature (the number in the parenthesis is the number of the database).[28,38] The list of chemical formulas and their ionic conductivity values are provided in **SI as a separate CSV file**. The distribution of ionic conductivity is shown in **Figure 2(a)**. Based on the distribution, only one material ($Na_{3.4}Mg_{0.4}Cr_{1.6}(PO_4)_3$) has ionic conductivity similar to that of a liquid electrolyte ($\sigma = 10^{-2}$ S/cm). 133 materials (50.4%) are in the $10^{-4}$ S/cm $\leq \sigma < 10^{-2}$ S/cm such as $Li_{1.1}Zr_{1.9}Y_{0.1}(PO_4)_3$ and $Li_{1.15}Zr_{1.85}Y_{0.15}(PO_4)_3$. Also, there are 78 materials (29.5%) in the $10^{-6}$ S/cm $\leq \sigma < 10^{-4}$ S/cm and 52 materials (19.7%) materials in the $\sigma < 10^{-6}$ S/cm. This means that it is challenging to find SSE materials exhibiting liquid electrolyte-like conductivity.

For the construction of classification model, we distinguish materials as superionic conductors and non-superionic conductors based on the decision boundary of $\sigma = 10^{-4}$ S/cm.[7,47] According to this decision boundary, the training database consisted of 134 superionic and 130 non-superionic conductors (**Figure 2(b)**).

To find superionic Li-SSEs, Li-containing materials (20,237) are obtained from MP. The screening process proceeds to distinguish Li-SSE candidates with a high likelihood of superionic nature. As a result, 468 Li-SSE candidates with potential superionicity. The type of 468 Li-SSE candidates is shown in **Figure 2(c)**. It shows Li-SSE candidates contain 299 oxide (63.9%), 96 halide (20.5%), 35 nitride (7.48%), 25 sulfide (5.34%), 7 selenide (1.50%), 3 hydride (0.64%), 2 phosphide (0.43%) and 1 carbide (0.21%) type. We note that as expected, the majority of materials are oxide, halide, and nitride because the oxide type NASICON, LISICON, garnet, perovskite, sulfide type argyrodite, nitride type, hydride type, and halide type were studied extensively as SSEs.[3,18,19,22,23] It is also important to note that the materials' chemical space in those types is larger than the others; hence there is still a great chance to find promising SSEs. In addition, there is a possibility of finding relatively unexplored materials types that have not been researched in details.

**Feature Engineering**

The prediction accuracy of the surrogate model is mainly dependent on the selection of appropriate features. From the previous work, for example, 127 NASICON and LISICON materials with 49 features consisting of simple molecular descriptors, structural descriptors, and electronic descriptors, 211 NASICON materials with Sure Independence Screening (SIS) and machine-learned ranking (MLR) are considered.[28,38] When both databases are combined, the largest number can be obtained, but the types of features available for each database are inevitably different. Therefore, the following study is conducted to use a feature that can apply both databases with using as many databases as possible.

In general, using CD applies to all databases, but it is impossible to explain the detailed and more specific structural attributes. Thus, to identify that this problem can cause decreasing the prediction accuracy, the results according to the composition of the features are considered. In addition, to confirm that the prediction accuracy was improved according to the database size, the results of two different database sizes are considered. The performances of the nine algorithms are compared to consider which models are more proper for ensembles and relatively more sensitive to features. For cross-validation, 100 different train-validation sets are considered. In addition, hyperparameter optimization is performed for each algorithm.

First, to consider the prediction accuracy dependency with respect to the composition of features, (1) CD and (2) the composition of CD and additional structural properties are considered. It is important to note that the CD can be easily generated using chemical formula but the containing information could be too general to describe the underlining physicochemical features. Additional structural properties contain detailed and more specific structural attributes, however, additional calculations or experiments could be necessary for construction. Mean accuracy (ACC) and standard deviation (STD) values of the prediction accuracy when the different composition of features (CD, CD + additional structural properties) is implemented were compared as shown in **Table S2(a), S2(b), and SI**. Comparing two different compositions, only 0.22% difference is shown for ACC and 0.24% difference for STD. Such results indicate that although only CD is used, it shows as much accuracy as it considers specific materials' structural properties. For easy access to a wide range of materials, only CD is used. Second, to compare the size of database, 123 and 264 sizes are considered (**Table S2(a), S2(c), and, SI**). As the size of the database increased to 264, the ACC increased by 7.87%, and STD decreased by 2.48%. Finally, the training database consisted of 264 materials and CD.

**Model Construction**

Comparing nine algorithms trained the training database size 264 and CD, DT, GB, and LGBM model performed the best. The overall mean accuracy is shown in **Figure S1, SI**, and detailed information is shown **Table S2(c), SI**. The performance of the DT model is the highest as 0.852 ACC and 0.043 STD, but there is a concern of overfitting, thus models showing the next performance were used. In addition, since a boosting model decreases the bias and variance,[48] tree-based boosting models GB, LGBM models are applied. Each GB, LGBM model performed 0.843 and 0.829 ACC and 0.042 and 0.048 STD, respectively (**Figure 3(a) and 3(b)**). GB and LGBM models are both used to construct a surrogate model because it is to utilize the performance of both models through ensembles rather than fully trusting only one model. The confusion matrices are shown in **Figure 3(c).** In addition, hyperparameter and top 20 feature importance of the GB, LGBM models are in **Table S3, S4, and SI**. The top 20 feature importance is obtained the average of the feature importance of each model trained with 100 different train-validation sets. Feature importance of the GB model is impurity-based, and the sum of importance of each feature is 1. Dev_SpaceGroupNumber has importance of 23.9%, and the remaining features are less than 10%. For the LGBM model, feature importance is calculated by numbers of times the feature is used in a model. The feature most frequently used in the LGBM model is dev_Gsvolume_pa, which is used an average of 9.3 times, followed by frac_dValence, Comp_L2Norm, and mean_GSvolume_pa about 7 times. The intersection of the feature importance of the GB, LGBM model is shown in **Table S5, SI**. There is a difference in the method of obtaining the feature importance in each GB, LGBM model, but 11 features are mainly used in both models. The 11 features consist of 10 elemental property-based attributes and one stoichiometric attribute.

**Prediction of the Superionic Li-SSE candidates**

Using a surrogate model ensembling GB and LGBM model, the ionic conductivity class of 468 Li-SSE candidates is predicted. In addition, to obtain the statistically meaningful results, the compositions of 100 different train-validation sets with different random states are used for prediction. Then, we considered the number of times predicted as a superionic conductor as shown in **Figure 4**. It indicates that the prediction for all 100 results to be superionic is 72 from the GB model and 118 from the LGBM model. For reducing the prediction uncertainties in each trained model, we intersected the predicted results of each model. As a result, 42 superionic Li-SSE candidates are suggested. A list of 42 predicted superionic Li-SSE candidates were shown in **Table S6, SI**. We note that 42 candidates consisted of oxide (39) and halide type (3). Again, the oxide materials comprise the majority (93%) of the suggested structures.

**Validation with AIMD**

For confirmation of predicted 42 superionic Li-SSE candidates, AIMD calculations are performed to compute the ionic conductivity at RT. In addition, since the ionic conductivity calculated through AIMD calculations and what was obtained through experiments are almost identical, the materials we calculated with AIMD calculations have been verified.[49–51] It is unfortunate that 4 materials have convergence issues during AIMD simulations, so omitted. As a result, six materials exceeding the decision boundary ($\sigma = 10^{-4}$ S/cm), and seven materials are close ($\sigma \geq 10^{-5}$ S/cm) to the boundary. MP id, chemical formula, and ionic conductivity at RT of these 13 materials are provided in **Table 1.** It is important to note that seven of the 13 materials were already studied as SSEs although their ionic conductivity values are not obtained yet, especially $LiTmSiO_4$, $LiInSiO_4$ as LISICON, and $Li_7La_3(SnO_6)_2$, $Li_3Nd_3(TeO_6)_2$ as garnet.[52–56] We also note that $K_3Ba_3Li_2Al_4B_6O_{20}F$, $Sr_3LiSbO_6$ were studied as nonlinear optical (NLO) materials and LED, not SSE, respectively.[57,58] Seven materials did not exceed the decision boundary ($\sigma = 10^{-4}$ S/cm) we set, but among them, many of the materials studied as SSE were distributed, so a wider boundary was set. Thus, we propose six new superionic Li-SSE candidates that have not been previously studied as SSE; $Cs_2LiNd(BO_3)_2$, $NaLiB_4O_7$, $Li_3Pr_3(TeO_6)_2$, $K_3Ba_3Li_2Al_4B_6O_{20}F$, $LiNd_6B_3O_{14}$, and $Sr_3LiSbO_6$.

## CONCLUSIONS

In this study, we constructed a platform consisting of a screening process and machine learning model to find new superionic Li-SSE candidates. Using the database consisting of ionic conductivity of NASICON, LISICON type SSE, and previously developed CD, a surrogate model was constructed to predict superionic conductors. 9 algorithms were considered to find optimal models for the ensemble. 100 different train-validation sets are considered for cross-validation. Ensembling of GB and LGBM models, the surrogate model was constructed. Each model performed 0.843, 0.829 ACC and 0.042, 0.048 STD, respectively. Through the screening process, 468 Li-SSE candidates with high superionic potential were derived from 20,237 Li-containing materials. Using the surrogate model, the ionic conductivity class of 468 superionic Li-SSE candidates at RT was predicted. 42 predicted superionic Li-SSE candidates are derived. To confirm the ionic conductivity of these 42 candidates, AIMD calculations proceeded. As a result, 13 superionic and semi-superionic Li-SSE candidates are identified. Excluding 7 materials previously studied as SSE, we suggest new 6 superionic Li-SSE candidates such as $Cs_2LiNd(BO_3)_2$, $NaLiB_4O_7$, $Li_3Pr_3(TeO_6)_2$, $K_3Ba_3Li_2Al_4B_6O_{20}F$, $LiNd_6B_3O_{14}$, and $Sr_3LiSbO_6$. We expect the constructed platform to accelerate finding new superionic Li-SSE materials.

## ACKNOWLEDGMENTS

This work was supported by the National Research Foundation of Korea (NRF) grant funded by the Korea government (MSIT) (No. 2020R1F1A1066519). This work was supported by the National Supercomputing Center with supercomputing resources including technical support (KSC-2021-CRE-0293).

## APPENDIX A. SUPPORTING INFORMATION

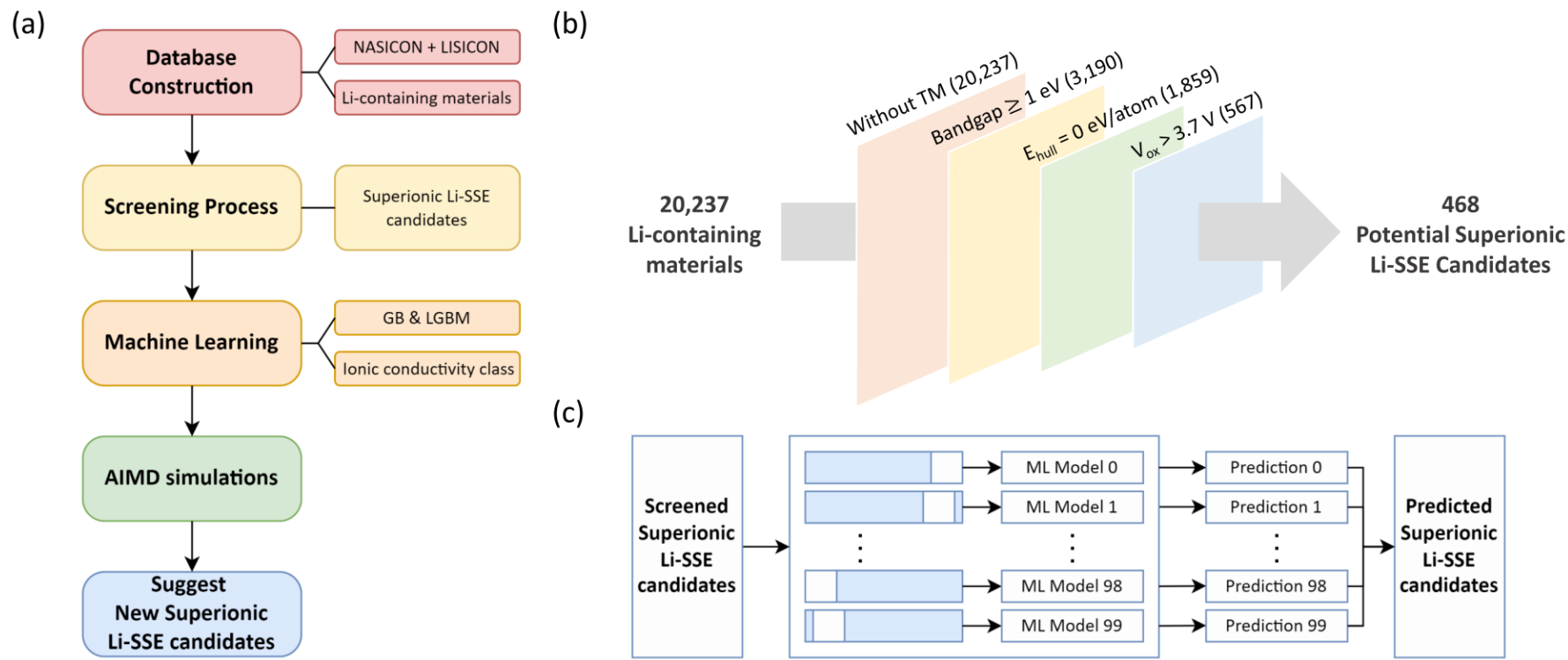

**Figure 1.** (a) A platform consisting of a screening process and machine learning model to find new superionic Li-SSE candidates. (b) Screening process to select Li-SSE candidates whose have high likelihood of superionic character based on atomic structural characteristic. (c) Machine learning process for predicting superionic Li-SSE candidates.

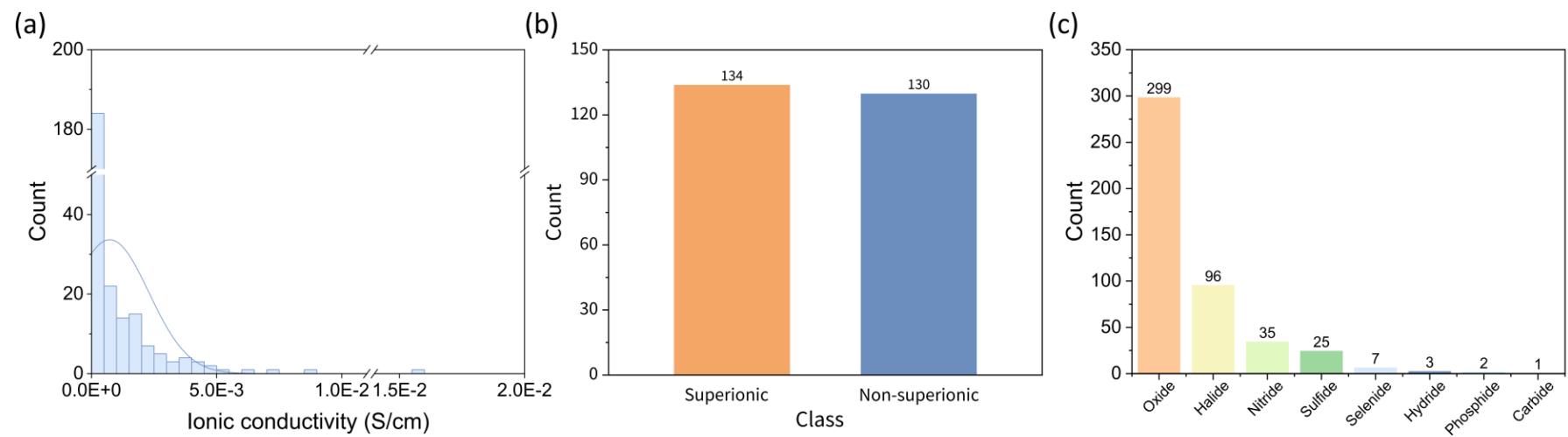

**Figure 2.** (a) Distribution of ionic conductivity of training database, (b) the ionic conductivity class of the training database that divided by σ = 10$^{-4}$ S/cm, (c) type of 468 Li-SSE candidates that have a high likelihood of superionic conductor.

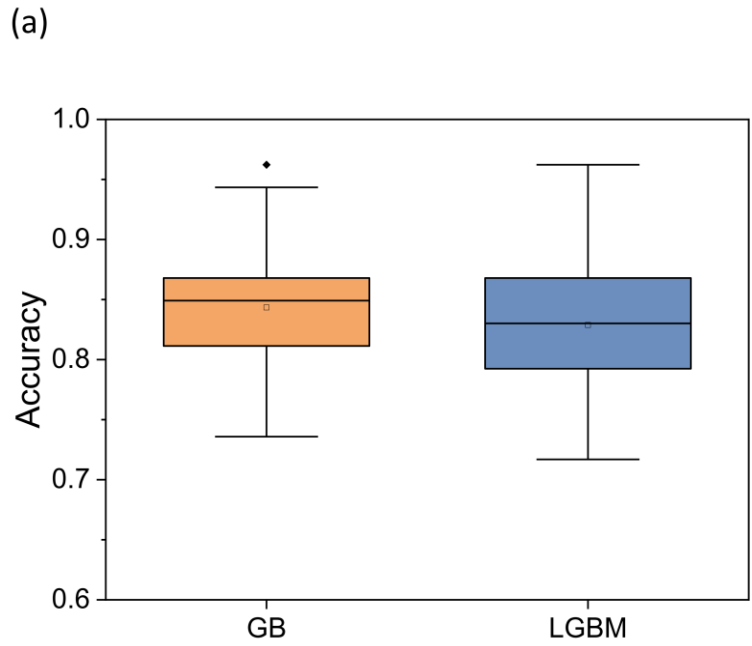
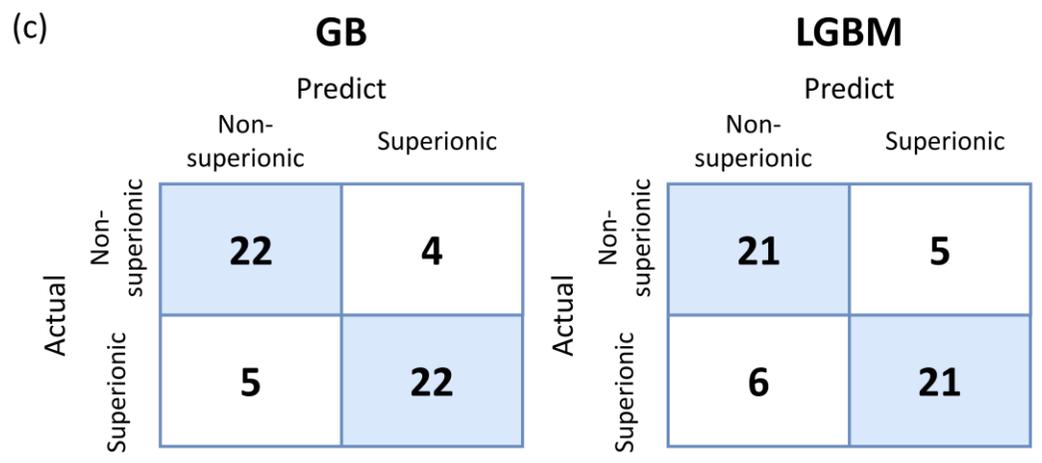

**Figure 3.** The surrogate model consisted of an ensemble of GB and LGBM models. Each GB, LGBM model has following performances; (a) distribution of ACC, (b) ACC and STD, and (c) confusion matrix, respectively.

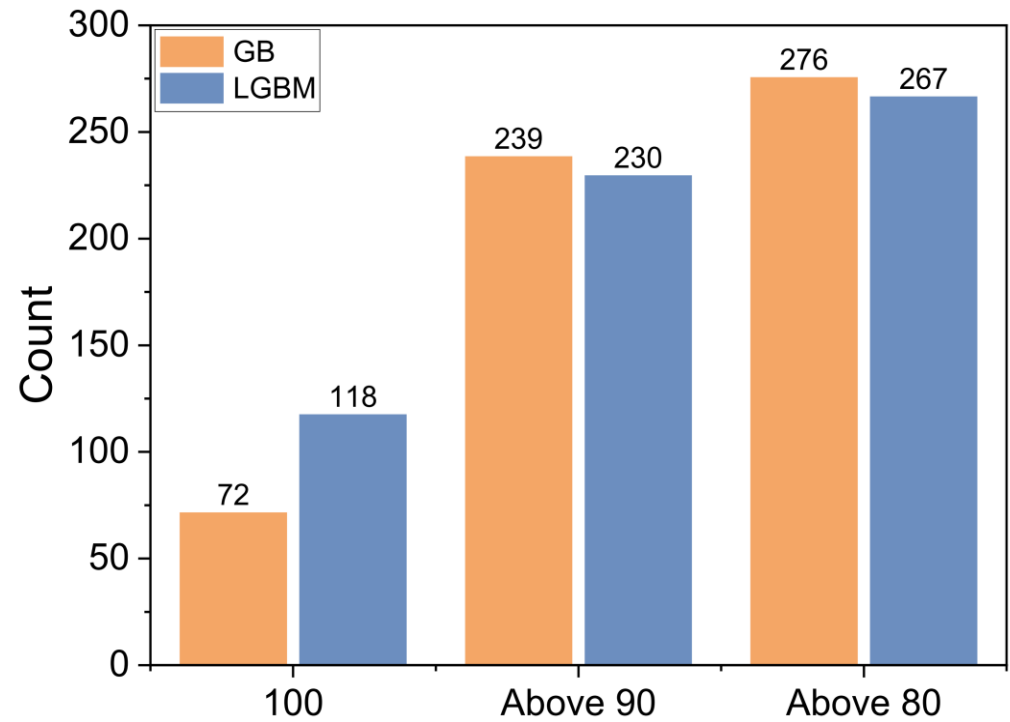

**Figure 4.** To obtain the statistically meaningful results, the compositions of 100 different train-validation sets with different random states are used for prediction. The number of times predicted as a superionic conductor for each GB, LGBM model.

| MP id | Chemical formula | Ionic conductivity (S/cm) at RT |
| --- | --- | --- |
| mp-1194993 | $Cs_2LiNd(BO_3)_2$ | 0.000503 |
| mp-18220 | $LiAlSiO_4$ | 0.000464 |
| mp-553926 | $NaLiB_4O_7$ | 0.000416 |
| mp-1211386 | $Li_3Pr_3(TeO_6)_2$ | 0.000381 |
| mp-4779 | $Li_2B_4O_7$ | 0.000341 |
| mp-754534 | $Li_2CeO_3$ | 0.0002 |
| mp-1195057 | $K_3Ba_3Li_2Al_4B_6O_{20}F$ | 8.60E-05 |
| mp-1211241 | $LiNd_6B_3O_{14}$ | 4.67E-05 |
| mp-1190151 | $Sr_3LiSbO_6$ | 3.68E-05 |
| mp-1200057 | $Li_7La_3(SnO_6)_2$ | 3.64E-05 |
| mp-15066 | $LiTmSiO_4$ | 2.61E-05 |
| mp-7205 | $LiInSiO_4$ | 2.34E-05 |
| mp-677627 | $Li_3Nd_3(TeO_6)_2$ | 1.79E-05 |

**Table 1.** MP id, chemical formula and ionic conductivity at RT for predicted 42 superionic Li-SSE candidates are shown. For confirmation of predicted 42 superionic Li-SSE candidates, AIMD calculations are performed to compute the ionic conductivity at RT.